\documentclass[aps,prl,floatfix,twocolumn,amsmath,amssymb]{revtex4-1}
\usepackage                             {epsfig}
\usepackage                             {xspace}
\usepackage                             {amsmath}
\usepackage                             {graphicx}
\usepackage     [english]               {babel}
\usepackage                             {natbib}
\usepackage                             {hyperref}
\usepackage                             {amsfonts}
\usepackage                             {amssymb}
\usepackage                             {latexsym}
\usepackage		                {color}
\usepackage       			{ulem}
\newcommand{\figref}[1]{Fig.~\ref{#1}}

\newcommand{\eref}[1]{(\ref{#1})}
\renewcommand{\eqref}[1]{equation (\ref{#1})}
\newcommand{\ket}[1]{| #1 \rangle}

\newcommand{\braket}[1]{\langle #1 \rangle}

\begin{document}

\title{Ultracold-atom quantum simulator for attosecond science}

\author{Simon Sala, Johann F\"orster, and Alejandro Saenz}

\affiliation{AG Moderne Optik, Institut f\"ur Physik,
  Humboldt-Universit\"at zu Berlin, Newtonstra{\ss}e 15, 12489 Berlin,
  Germany}

\date{\today}
\begin{abstract}
  A quantum simulator based on ultracold optically trapped atoms for
  simulating the physics of atoms and molecules in ultrashort intense
  laser fields is introduced. The slowing down by about 13 orders of
  magnitude allows to watch in slow motion the tunneling and
  recollision processes that form the heart of attosecond science. The
  extreme flexibility of the simulator promises a deeper understanding
  of strong-field physics, especially for many-body systems beyond the
  reach of classical computers. The quantum simulator can
  experimentally straightforwardly be realized and is shown to recover
  the ionization characteristics of atoms in the different regimes of
  laser-matter interaction.
\end{abstract}

\maketitle



In his renowned lecture, ``Simulating physics with computers''
\cite{qs_extra:feyn82} Richard P.\ Feynman suggested the use of quantum
simulators, i.e.\ precisely controllable quantum systems, to simulate other
quantum systems that cannot be described theoretically due to their
exponentially growing Hilbert space. For instance, the Mott-insulator to
superfluid phase transition in condensed-matter systems \cite{qs_extra:fish89}
was predicted \cite{cold:jaks98} to be observable with ultracold atoms in an
optical lattice and then successfully demonstrated
\cite{cold:grei02a,qs_extra:stof04}. Also the Higgs mechanism
\cite{qs_extra:endr12}, high temperature superconductivity
\cite{qs_extra:schn08}, or \textit{Zitterbewegung} \cite{cold:gerr10} (to name
just a few) were successfully investigated by quantum simulation. Moreover,
the quantum simulation of electrons in crystalline solids exposed to laser
fields \cite{cold:arli10} has been proposed.

\begin{figure}[ht]
  \begin{centering}
    \includegraphics[width=0.5\textwidth]{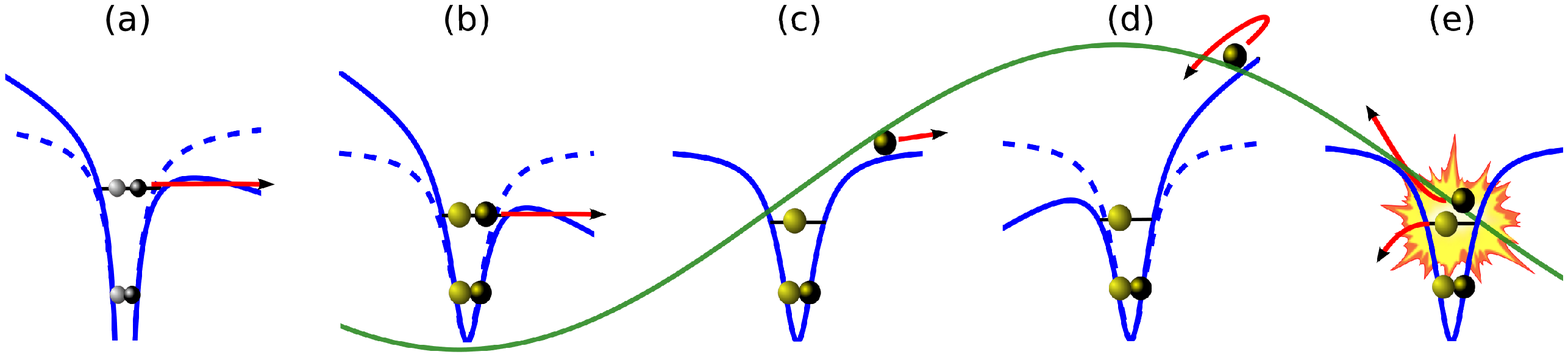}
    \caption{(color online) \textbf{(a) \& (b)}: Comparison of
      electrons in an atom exposed to a strong electric field {(a)}
      and atoms in an optical trap exposed to a magnetic-field
      gradient {(b)}. The different shadings of the electrons and
      atoms reflects their different spin states and Zeeman substates,
      respectively.  An external electric field {(a)} or
      magnetic-field gradient {(b)} effectively tilts the continuum
      threshold and the electrons {(a)} or atoms {(b)} can escape the
      binding potential by
      tunneling.\\
      \textbf{(b) - (e)}: Behavior of optically trapped atoms in a
      periodically driven magnetic-field gradient (solid green curve),
      as expected from the three-step model \cite{sfm:cork07} in
      strong-field physics. After tunneling {(b)} the escaped atom
      accelerates {(c)}, reverses {(d)} and finally recollides {(e)}
      with the residual atoms.  }
    \label{fig:fig1}
  \end{centering}
\end{figure}

Strong-field physics has contributed considerably to the understanding
of the light-matter interaction. The progress leading to pulses on the
attosecond timescale \cite{sfa:paul01a} has even raised visions of
real-time imaging of molecular processes \cite{sfm:haes10} and orbital
tomography \cite{sfm:itat04}.  Yet, attosecond many-body physics is
challenging. An exact investigation on classical computers beyond the
single-active-electron approximation becomes prohibitively complex for
many-electron systems. In fact, the numerical treatment of
two-electron systems like He or H$_2$ is today still state of the art
\cite{sfm:vann10, sfm:dehg10,sfa:arms11,sfm:silv12}. Thus, simplified
models are widely used for interpreting modern experiments. These
models are controversial and their validation is difficult for several
reasons. First, the used light pulses are bound to the specifications
of the laser. The wavelength range of lasers is limited, mostly
Ti:sapphire lasers are used. The pulse shapes are restricted and can
often only be reproduced and determined up to a considerable
uncertainty. The intensity and timescale of laser pulses are already
pushed to a limit where further improvements require major technical
or even principle developments with new limitations, like
free-electron lasers. Second, atoms, ions, and molecules are
complicated many-body systems. Their internal structure cannot be
simply manipulated. For example, a variation of the number of
electrons or protons underlies constraints due to electroneutrality.
Third, although the correlation of electronic and nuclear motion is
known to influence the ionization behavior \cite{sfm:pala06,
  sfm:silv13}, in most theoretical models this effect is neglected by
fixing the nuclei in space while investigating the electronic response
to the laser field.

In this work, we introduce the concept of an ultracold-atom quantum
simulator for attosecond science which offers great flexibility and
control beyond the mentioned limitations.  This includes many-body
quantum simulations that are impossible with any classical computer.

\textit{The attoscience simulator.} The simulator system consists of
ultracold trapped atoms that replace the electrons in the atom, ion,
or molecule, see \figref{fig:fig1}. The core potential is replaced by
an external, optical trapping potential.  The ability to implement
single-well or multi-well trapping potentials allows for a simulation
of atoms or molecules, respectively.  Naturally, fermionic atoms may
be chosen, but using bosons or distinguishable particles reveals
effects of the exchange interaction. The intense laser pulse is
replaced by a periodically driven magnetic-field gradient which is
generated by current-carrying coils.  Restrictions for ultrashort
laser fields like the zero-net-force condition \cite{sfa:milo06} do
not apply here and thus fields of almost arbitrary shape can be
created, even true half-cycle pulses and fields that formally
correspond to sub-attosecond pulses.

Certainly, the atom-atom interaction is shorter ranged than the
Coulomb interaction. However, earlier quantum simulations like the
famous superfluid to Mott-insulator phase transition
\cite{cold:jaks98} demonstrated that an equivalent physics is
obtainable. The use of ultracold atoms introduces the unique
opportunity to arbitrarily vary the effective interaction strength
{\it via} magnetic Feshbach resonances. This promises new insights on
the influence of the interparticle interaction on the ionization
behavior. Furthermore, theoretical studies which replace the core
potential by, e.g., a zero-range potential \cite{sfa:milo06}, can now
be tested experimentally, and this even for many-particle systems.
Since ultracold quantum systems are manipulated nowadays on the
single-atom level \cite{cold:serw11,cold:weit11}, important tests of
the widely used single-active-electron approximation and a detailed
investigation of correlated many-body tunneling become
accessible. Moreover, only the simulator allows for the experimental
realization of fixed nuclei -- a task impossible with real molecules
due to the Heisenberg uncertainty principle.  The influence of a fixed
nuclear geometry on the ionization behavior \cite{sfm:saen00c} can
thus be tested experimentally in a clean fashion. Additionally, the
differences between the quantum-mechanical nature of vibronic states
and the simulation of a mechanical vibration of the nuclei can be
investigated.


\textit{Hamiltonian mapping.} The formal equivalence of the quantum
simulator Hamiltonian to the electronic strong-field Hamiltonian at a
fixed nuclear configuration is demonstrated. When treating the strong
laser field classically, which is acceptable due to its high
intensity, and applying dipole approximation and length gauge (LG),
respectively, the electronic strong-field Hamiltonian reads
\begin{align}
  \label{eq:sf_hamil}
  \hat{H}^{\mathrm{LG}}(t) = \hat{H}_0 + \sum_{i=1}^{N} \mathbf{r}_i \cdot  e \mathbf{E}(t) \quad ,
\end{align}
where
\begin{align}
  \label{eq:sf_free_hamil}
  \hat{H}_0 = \sum_{i=1}^{N} \frac{\mathbf{\hat{p}}_i^2}{2 m_{\mathrm{e}}} +
  V_{\mathrm{ee}} + V_{\mathrm{e,nuc}}
\end{align}
denotes the field-free Hamiltonian for $N$ electrons. $m_{\mathrm{e}}$ is the
electron mass, $e$ the electron charge, $V_{\mathrm{ee}}$ includes all
electron-electron repulsion terms, and $V_{\mathrm{e,nuc}}$ all the
electron-nucleus interactions. $\mathbf{E}$ denotes the electric-field
component of the pulse. In analogy, the Hamiltonian of $N$ ultracold atoms
confined in a trapping potential $\mathcal{V}_{\mathrm{a,tr}}$ which are
exposed to a time-dependent magnetic-field gradient
$\boldsymbol{\mathcal{B}}^{\prime}(t)$ reads
\begin{align}
  \label{eq:uc_hamil}
  \hat{\mathcal{H}}^{\mathrm{LG}}(t)= \hat{\mathcal{H}}_0  + \sum_{i=1}^{N} \mathbf{r}_i \cdot \mu
  \boldsymbol{\mathcal{B}}^{\prime}(t) \quad ,
\end{align}
where 
\begin{align}
  \label{eq:uc_free_hamil}
 \hat{\mathcal{H}}_0 =  \sum_{i=1}^{N} \frac{\mathbf{\hat{p}}_i^2}{2 m_{\mathrm{a}}} +
  \mathcal{V}_{\mathrm{aa}} + \mathcal{V}_{\mathrm{a,tr}} 
\end{align}
denotes the Hamiltonian of the atoms in the trap without the gradient
$\boldsymbol{\mathcal{B}}^{\prime}$. $m_{\mathrm{a}}$ denotes the
atomic mass, $\mu$ the magnetic moment of the atoms,
$\mathcal{V}_{\mathrm{aa}}$ includes all atom-atom and
$\mathcal{V}_{\mathrm{a,tr}}$ all atom-trap interactions,
respectively.
 
The Hamiltonians \eref{eq:sf_hamil} and \eref{eq:uc_hamil} are
\textit{formally} equivalent under the mapping
\begin{align}
  \label{eq:mapping}
  e \ \mathbf{E}\, \mapsto\, \mu \ \boldsymbol{\mathcal{B}}^{\prime} \quad .
\end{align}
It is important to note that the electrodynamical potentials,
e.g., the vector potential $ -\frac{\partial \mathbf{A}(t)}{\partial t}
= \mathbf{E}$, map accordingly. In the ultracold simulator system
the ``vector potential'' is thus given by
\begin{align}
  \label{eq:vector_pot}
  -\frac{\partial \boldsymbol{\mathcal{A}}(t)}{\partial t} =
  \mathbf{\boldsymbol{\mathcal{B}}^{\prime}} \quad .
\end{align}
Of course, the potential $\boldsymbol{\mathcal{A}}$ differs from the
physical vector potential $\mathbf{\tilde{A}}$ that generates the
magnetic field $\boldsymbol{\mathcal{B}}$ and its gradient
$\boldsymbol{\mathcal{B}}^{\prime}$ via $\boldsymbol{\mathcal{B}} =
\nabla \times \mathbf{\tilde{A}}$. Yet, equation \eref{eq:vector_pot}
is the formal consequence of the simulator mapping \eref{eq:mapping}.

The simulator mapping \eref{eq:mapping} is intrinsically defined in length
gauge. However, it is particularly useful to consider the analog of the
velocity-gauge (VG) formulation, too.
A gauge transformation of the strong-field Hamiltonian \eref{eq:sf_hamil}
leads to the velocity-gauge form
\begin{align}
  \hat{H}^{\mathrm{VG}}(t)= \hat{H}_0 + \sum_{i=1}^{N} \frac{e}{m_{\mathrm{e}}}\
  \mathbf{A}(t) \cdot \mathbf{\hat{p}}_i  +\frac{e^2}{2m_{\mathrm{e}}}\
   \mathbf{A}(t)^2 
   \label{eq:tdsevg_sf}
\end{align}
In analogy, a ``gauge'' transformation of the simulator Hamiltonian
\eref{eq:uc_hamil} leads to the corresponding simulator Hamiltonian
in ``velocity gauge'',
\begin{align}
  \hat{\mathcal{H}}^{\mathrm{VG}}(t)= \hat{\mathcal{H}}_0 + \sum_{i=1}^{N} \frac{\mu}{m_{\mathrm{a}}}\
  \boldsymbol{\mathcal{A}}(t) \cdot \mathbf{\hat{p}}_i +\frac{\mu^2}{2m_{\mathrm{a}}}\
  \boldsymbol{\mathcal{A}}(t)^2 
  \quad .
  \label{eq:tdsevg_uc}
\end{align}
Again, the Hamiltonians are \textit{formally} equivalent. The vector
potential $\boldsymbol{\mathcal{A}}(t)$ is the one resulting from the
simulator mappings
\eref{eq:mapping} and \eref{eq:vector_pot}.\\

\textit{Experimental realization.} As a possible realization we
consider the experiment of the group of Selim Jochim
\cite{cold:serw11} where a well defined number of fermionic atoms can
be loaded into a tight optical dipole trap in a well defined quantum
state. This trap is in good approximation described by a
one-dimensional Lorentz potential \cite{qs_extra:sm}. A
\textit{static} magnetic-field gradient which tilts the continuum
threshold of the trap, see \figref{fig:fig1}, is applied for the
preparation and investigation of the system. The here proposed
attoscience quantum simulator is realized by replacing the static
magnetic-field gradient by a periodically driven one.

Similarly to strong-field physics, the pulse may be defined by its
vector potential {\it via}
\begin{align}
  \boldsymbol{\mathcal{A}}(t) = \boldsymbol{\mathcal{A}}_0 \, \sin\left(
    \frac{\omega t}{2n_{\mathrm{c}}} \right)^2\, \sin(\omega t + \varphi)
  \quad .
\end{align}
Here, $n_{\mathrm{c}}$ is the number of cycles, $\varphi = 0$ is the
carrier-envelope phase, $\omega$ the angular frequency, and
$\boldsymbol{\mathcal{A}}_0$ from which
$\boldsymbol{\mathcal{B}}_0^{\prime}$ is obtained {\it via} equation
(\ref{eq:vector_pot}) is the strength of the perturbation.  For a given
simulator setup and a specific pulse, the corresponding values for the
frequency $\omega_{\mathrm{e}}$ and peak vector potential
$|\mathbf{A}_0|$ applied in the strong-field system are found by
enforcing equal Keldysh parameters \cite{sfa:keld65,sfm:cork07} and an
equal ratio of the binding energy to the frequency of the perturbing
field \cite{qs_extra:sm}.

\begin{figure}[ht]
  \begin{centering}
    \includegraphics[width=0.48\textwidth]{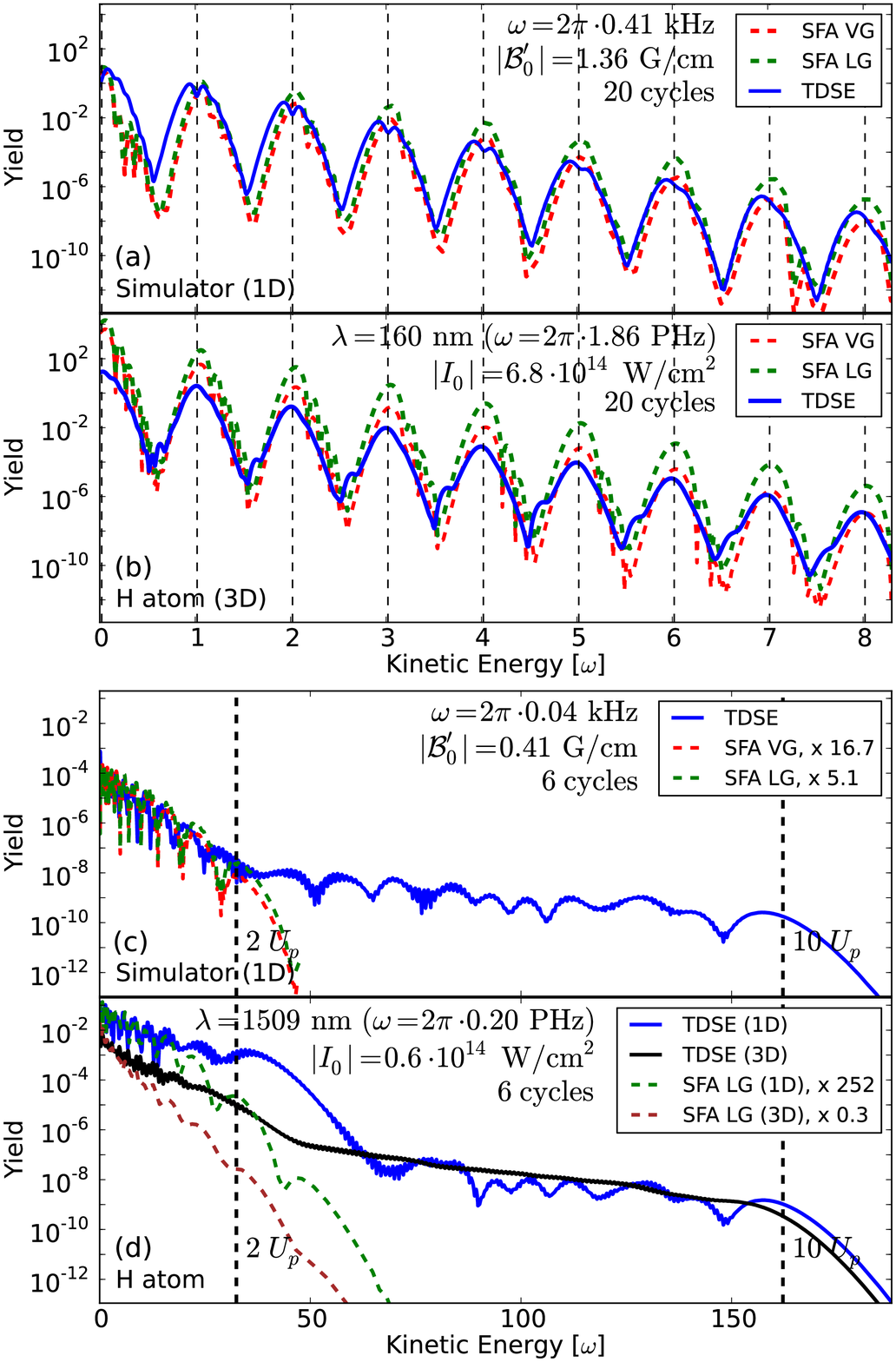}
    \caption{(color online) Atom spectra of the simulator and electron
      spectra of a hydrogen atom for the multiphoton regime, (a) and
      (b) respectively, and for the quasistatic regime, (c) and (d)
      respectively. The dashed vertical lines in (a) and (b) indicate
      the positions of the multiphoton peaks for an infinitely long
      pulse as expected from the subsequent absorption of field
      quanta.  In (d), in addition to the result for the hydrogen atom
      in three dimensions, also the result for the one-dimensional
      (1D) \textit{soft-Coulomb} potential $V(z)=-1/\sqrt{2+z^2}$ is
      shown.  The SFA yields in (c) and (d) are rescaled in order to
      agree with the total ionization yield of the TDSE
      calculation. The factors are given in the figure legends.  }
    \label{fig:fig2}
  \end{centering}
\end{figure}

\textit{Validation of the quantum simulator.}  In the experiment
\cite{cold:serw11} the simulator builds on, the atom loss is routinely
measured. This observable corresponds to a measurement of the total
ion (or electron) yield in a strong-field experiment.  More detailed
information on the underlying physics is obtained by a measurement of
differential yields: energy-resolved electron or atom spectra for
strong-field experiments or the simulator, respectively. The
measurement of energy-resolved atom spectra requires further
experimental developments, similarly to strong-field physics where in
the early days also only total yields were measured. To validate the
simulator in more detail, energy-resolved electron spectra of a
hydrogen atom are compared to the corresponding energy-resolved atom
spectra of the simulator setup, both initially in their ground
state. The spectra are calculated by solving the corresponding
time-dependent Schr\"odinger equations (TDSE) \cite{qs_extra:sm},
ensuring that the corresponding parameters for the quantum simulator
are experimentally accessible \cite{qs_extra:selim_priv}.

The laser-matter interaction is typically divided into two
characteristic regimes. In the low-frequency, high-intensity regime
the system is assumed to follow adiabatically the changes of the
electric field of the laser. In this quasistatic regime the electron
is supposed to tunnel through or escape over the field-distorted
potential barrier, see \figref{fig:fig1}. In the other limit of the
high-frequency, low-intensity regime the multiphoton picture is
usually adopted in which the ionization is described within a
simplified picture as an absorption of photons, despite the fact that
in the theoretical treatment the electromagnetic field is treated
classically.

In the multiphoton regime (Figs.~\ref{fig:fig2}a and b), both spectra
show the typical multi-peak structure (above-threshold-ionization
peaks) where the peak distance reflects the frequency of the
perturbing field. Clearly, simulator and hydrogen atom show very good
agreement. Despite the different dimensionalities the TDSE solutions
agree, in fact, almost quantitatively.

In the quasistatic regime (Figs.~\ref{fig:fig2}c and d), a simple
tunneling picture suggests an exponential decrease in the
energy-resolved spectra, as is seen in the low-energy part (up to $2
U_p$ where $U_p = I /(4 \omega^2)$ is the ponderomotive energy and $I$
the laser intensity). However, in a periodically changing field the
emitted electron or atom can reverse its direction of motion and
recollide, see \figref{fig:fig1}. High-harmonic generation in
strong-field physics is based on the recombination of the liberated
electron with the parent ion at the recollision step. Using classical
Newtonian mechanics it had been found that high-harmonic spectra
extend up to $3.17U_p + I_p$ \cite{sfa:cork93}. For energy resolved
electron spectra, the recollision process leads to a broad energy
distribution of the rescattered electrons which manifests in a plateau
as observed in \cite{sfa:scha93} and clearly seen in
\figref{fig:fig2}d.
In analogy to the high-harmonic cutoff law classical Newtonian
mechanics predicts an extension of this plateau between $2 U_p$ and
$10 U_p$ \cite{sfa:paul94b}. Clearly, the simulator shows all expected
features, both from tunneling and rescattering. However, the more
pronounced structures in the plateaus of the 1D systems (simulator and
1D hydrogen atom) reveal effects of the dimensionality. Such effects
can be studied with the simulator even experimentally by varying the
anisotropy of the trap -- a task impossible in strong-field
experiments.

Rescattering is the origin of nonsequential double ionization,
high-energy above-threshold ionization, and high-order harmonic
generation. A controlled recollision, see \figref{fig:fig1}, of an
escaped atom on residual bound atoms prepared in a specific
configuration with variable interaction strength can reveal insights
into correlated recollision dynamics relevant, e.\,g., for high
harmonics \cite{sfa:shaf12} and non-sequential double ionization
\cite{sfa:berg12}. On the other hand, inspired by the experiments on
imaging molecular orbitals using laser-induced electron tunneling and
diffraction \cite{sfm:meck08} controlled rescattering collisions can
serve for the imaging of ultracold many-body wavefunctions.

\textit{Strong-field approximation.} In the widely used strong-field
approximation (SFA) \cite{sfa:keld65,sfa:fais73,sfa:reis80} bound
states of the potential other than the initial state are neglected and
the final continuum state is replaced by a Volkov state, i.e.\ the
solution of a free electron in a laser field. Therefore, the
interaction of the electron with the remaining ion is ignored in the
final state. Thus, the SFA does not support rescattering as can also
be seen in Figs.~\ref{fig:fig2}c and d.
On the other hand, the direct electrons and atoms (up to $2 U_p$) in
the quasistatic regime in Figs.~\ref{fig:fig2}c and d are
qualitatively well described by the SFA. Similarly, the SFA reproduces
the multi-peak structure in the multiphoton regime, see
Figs.~\ref{fig:fig2}a and b. 
Note, the SFA is not gauge invariant and so far no arguments from
first principles are known what gauge is to be preferred in which
situation \cite{sfa:vann09} (the gauge problem of SFA). In the
simulator system, the number of trap states as well as the potential
range can be varied which allows for an analysis of the assumptions
and the gauge ambiguity of the SFA.

Interestingly, the SFA in velocity gauge allows to obtain the momentum
density of the initial state since the energy-resolved yield is a
product of the momentum-space density $|\tilde{\psi}(\mathbf{p})|^2$
and a prefactor $|g(\mathbf{p})|^2$ \cite{qs_extra:sm}. For a given
momentum $\mathbf{p}$, the prefactor $g(\mathbf{p})$ depends solely on
the vector potential $\boldsymbol{\mathcal{A}}(t)$ and the binding
energy. In contrast to the corresponding strong-field experiments,
these parameters are known precisely for the quantum simulator because
of the exactly known pulse shape. Note, this imaging technique relies
on the agreement of the SFA in velocity gauge with the full TDSE
results, which is fulfilled as seen in
Figs.~\ref{fig:fig2}a and c
despite the fact that the simulator mapping \eref{eq:mapping} is bound
to the length gauge. Thus, this imaging technique indeed allows to
image the momentum density of an ultracold gas in a trap.


\textit{Conclusion.} A proposal for a quantum simulator for attosecond
physics is presented based on ultracold atoms in an optical trapping
potential. The simulator idea connects the very contrary physics of
ultracold, trapped atomic gases and the one of atoms, ions, and
molecules in ultra-intense, ultra-short laser fields. The constraints
one faces in strong-field experiments, such as the limitation to a
specific molecular geometry, a fixed number of electrons per element
or molecule, fixed interaction strengths, and limited pulse shapes are
overcome in the simulator system. Moreover, the simulation can even
reach parameter regions which are beyond those nowadays realizable in
strong-field experiments, including, e.g., exotic pulse shapes and
effective pulse durations corresponding to the sub-attosecond regime.
In fact, the here proposed attosecond science in slow motion may shed
light onto the ongoing debate on tunneling times
\cite{sfa:uibe07,sfa:shaf12,sfa:pfei12,sfa:eckl08}. The numerical
analysis of the here proposed concrete experimental realization of the
quantum simulator with realistic experimental parameters demonstrates
that it reproduces in its simplest configuration the ionization
characteristics of a hydrogen atom. While this simple demonstrating
example can be evaluated computationally, the simulator paves the way
to systematically investigate many-body systems where the full
numerical treatment is beyond the reach of any classical computer.
Also the physics of ultracold atoms may profit from the quantum
simulator by adopting concepts developed in attosecond science.

\acknowledgments{The authors gratefully acknowledge Selim Jochim,
  Gerhard Z\"urn, Thomas Lompe and Andre N.\ Wenz for valuable discussions
  and details of their experiment, and financial support from the
  \textit{Studienstiftung des deutschen Volkes}, \textit{Fonds der
    Chemischen Industrie}, and the \textit{EU Initial Training Network
    (ITN) CORINF}.}


%

\clearpage
\newpage

\section{Supplemental Material}

\subsection{Experimental parameters}

The proposed experimental realization of the quantum simulator is based on the
extension of an existing experiment. In the experiment \cite{cold:serw11}, the
potential of a tight optical dipole trap is populated with a degenerate Fermi
gas consisting of $^6$Li atoms in two hyperfine states. By applying a
\textit{static} magnetic-field gradient and varying the trap depth the atoms
tunnel out of the trap in a fully controlled way, ending up with a defined
number of particles in a determined quantum state. The experimental dipole
trap is described in good approximation by a Gaussian-beam potential which
results in a quasi 1D confinement with an aspect ratio of about 1:10 . The
beam profile in the longitudinal direction is approximately given by the
Lorentz potential
\begin{align}
  \label{eq:v_l}
  \mathcal{V}_{L}(z) = \alpha \mathcal{V}_0 \left[1 -
    \frac{1}{1+(z/z_r)^2}\right] \quad .
\end{align}
Here, $\alpha$ is a modulation factor which allows to vary the trap
depth in the experiment, see \cite{cold:serw11, cold:zurn12},
$\mathcal{V}_0/k_b= 3.33\ \mathrm{\mu K}$ is the potential depth,
$k_b$ the Boltzmann constant, $z_r = \pi w_0^2 / \lambda$ the Rayleigh
length with a laser wavelength of $\lambda=1064$ nm.  In the
experiment, the value of the waist $w_0$ has been $1.8\ \mathrm{\mu
  m}$. Achieving values lower than this is challenging but in
principle a value of at least $w_0=0.7\ \mathrm{\mu m}$ could be
realized with a new experimental setup
\cite{qs_extra:serw_phd,qs_extra:selim_priv}. Here we chose $w_0=0.6\
\mathrm{\mu m}$.  Moreover, we chose a realistic value for the
trapping depth $\alpha = 0.02$. Certainly with the chosen Lorentz
potential the long-range interaction of the nuclei and the electrons
cannot be reproduced exactly. However, the choice of the parameters
are such that energetic distribution of the low-lying bound-states
resemble appropriately those of a 1D hydrogen atom with
\begin{eqnarray}
V(z) = - \frac{1}{\sqrt{2+z^2}} \quad .
\end{eqnarray}\\
On the other hand, systems with smaller binding potentials,
like, e.g., anions, could be simulated more accurately. For anions,
the long-range interaction of an emitted electron with the remaining
neutral atom scales as $1/r^4$. This is more comparable to the long range
interatomic interaction which scales like $1/r^6$ for neutral atoms or
with $1/r^3$ for dipolar atoms.

\subsection{Parameter mapping}

For the simulator system natural units (n.u.) are introduced in which
$\hbar$, the magnetic moment $\mu$, the atomic mass $m_{\mathrm{a}}$
and the trap length $d$, which is equal to the extension of the
ground-state wave function (defined as the distance where the
ground-state wavefunction has decreased to 1/e of its maximum value),
are set to unity. It should be emphasized that different to atomic
units (a.u.) which are uniquely defined to reflect the electronic
properties of the hydrogen atom, the introduced natural units change
with the trapping potential.

For a given simulator setup and a specific pulse, the corresponding
values for the frequency $\omega_{\mathrm{e}}$ and peak vector
potential $|\mathbf{A}_0|$ applied in the strong-field system are
found by enforcing equal Keldysh parameters
\begin{align}
  \gamma_{\mathrm{e}} := \omega_{\mathrm{e}} \frac{\sqrt{2 m_{\mathrm{e}}
      I_p}}{e E_0} = \omega \frac{\sqrt{2 m_{\mathrm{a}} E_{\mathrm{b}}}}{\mu
    {\mathcal{B}}^{\prime}_0} =: \gamma_{\mathrm{a}}
\end{align}
and equal parameters
\begin{align}
  \beta_{\mathrm{e}} := \frac{ I_p }{ \hbar \omega_{\mathrm{e}}} =
  \frac{E_{\mathrm{b}} }{ \hbar \omega } =: \beta_{\mathrm{a}} \quad .
\end{align}
Here, $I_p$ and $E_b$ are the binding energies of the ground state of the
field-free Hamiltonians (1) and (3) in the main manuscript,
respectively. This mapping is not unique since $E_b$, which is determined by
the shape of the trapping potential, is a free parameter. This freedom can be
used to better adjust the trapping potential such that the energy-level
distribution of the simulated system is resembled accurately.

In the manuscript, the considered pulses are defined via the vector
potential although the simulator mapping is performed in length gauge.
Since the simulator system is directly compared to the corresponding
strong-field system, this is a convenient way to ensure that the pulse
fulfills the zero net-force condition, i.\,e.\ the total integral over
its electric (or magnetic) field is zero and the vector potential has
the same value before and after the pulse.

\subsection{Solution of the TDSE}
The TDSE
\begin{align}
  i \frac{\partial}{\partial t} \ket{\psi(t)} = \hat{H} \ket{\psi(t)}
\end{align}
for the Hamiltonians $\hat{H}$ and $\hat{\mathcal{H}}$,
(1) and (3) in the main manuscript, respectively, are solved by
expanding the wavefunction $\ket{\psi(t)}$ in eigenstates $\ket{\phi}$
of the field-free Hamiltonians $\hat{H}_0$ and $\hat{\mathcal{H}}_0$,
(2) and (4) in the main manuscript, respectively.\\

While the finally shown spectra have been calculated within velocity gauge,
i.\,e.\ equations (7) and (8) in the main manuscript, to achieve faster
convergence, it was verified for a selected number of laser parameters that
length and velocity gauge results are in full agreement.

The simulator system is treated numerically in one dimension. This
approximation is well satisfied for the here considered experimental
realization since for ultracold temperatures the transversal motion of
a strongly anisotropic trapping potential as realized in
\cite{cold:serw11} is frozen out to the ground state.

For the solution of the 1D simulator [soft-Coulomb] system, the eigenstates
of the field-free Hamiltonian are calculated via the matrix algorithm
\cite{qs_extra:foer12}. The results shown in the manuscript were obtained with
a box size of $x_{\mathrm{max}}=1200$ n.u. [$x_{\mathrm{max}}=1200$ a.u.] and
a grid consisting of $4001$ [$4201$] points. For the time propagation, only
$3001$ [$2001$] states were used.\\

For the time propagation of the hydrogen atom, the field-free
eigenstates are obtained in spherical coordinates $(r,\theta,\varphi)$ by expanding
the radial problem in B splines (as in \cite{sfm:vann08}).
The results shown in the manuscript were obtained with a box size of 
$r_{\mathrm{max}}=2000$ a.u.\ including states with angular momenta up to $l=50$.
Along the radial coordinate $4000$ B splines of order $12$ and
a linear knot sequence were used.
For the time propagation, only states with energies up to $10$ 
a.u.\ above the ionisation threshold were considered.\\

Convergence of all spectra shown in the manuscript was ensured by varying the
basis.

\subsection{SFA in the simulator system}
In analogy to the TDSE, the direct SFA ionization amplitude
$\mathcal{M}_{\mathbf{p}}$ is easily translated from the corresponding
formula for the strong-field amplitude \cite{sfa:milo06} applying
equations (5) and (6) in the main manuscript. In length
gauge it reads
\begin{align}
  \mathcal{M}_{\mathbf{p}}^{\mathrm{LG}} &= -i \int_0^{t_f} dt\
  \braket{\psi^{V}(t) | \mathbf{r} \cdot
    \boldsymbol{\mathcal{B}}^{\prime}(t) | \phi_0(t) }\\
  &= \int_0^{t_f}
  dt\ e^{i S_{\mathbf{p}}(t)} \boldsymbol{\mathcal{B}}^{\prime}(t)
  \frac{d}{d\boldsymbol{\pi}(t)} \tilde{\phi}(\boldsymbol{\pi}(t))
  \quad ,
  \label{eq:sfalg}
\end{align}
where $\boldsymbol{\pi}(t) = \mathbf{p} +
\boldsymbol{\mathcal{A}}(t)$, $S_{\mathbf{p}}(t) = E_b\, t +
\frac{1}{2} \int_0^{t} d\tau \, \boldsymbol{\pi}(\tau)^2$ is the
classical action, $ \boldsymbol{\mathcal{A}}(t) $ is the vector
potential according to the mapping (6) in the main manuscript, 
$\tilde{\phi}$ is the Fourier transform of the initial state
wavefunction, and $\ket{\psi^{V}}$ is the Volkov wavefunction.
In velocity gauge, the amplitude is
\begin{align}
  \mathcal{M}_{\mathbf{p}}^{\mathrm{VG}} &= -i
  \tilde{\phi}(\mathbf{p}) \int_0^{t_f} dt\ e^{i S_{\mathbf{p}}(t)}
  \left( \mathbf{p} \cdot \boldsymbol{\mathcal{A}}(t) + \frac{1}{2}
    \boldsymbol{\mathcal{A}}(t)^2 \right)
  \label{eq:sfavg}
\end{align}
and thus the yield
\begin{align}
  \mathrm{Y} &= |\mathcal{M}_{\mathbf{p}}^{\mathrm{VG}}|^2 \equiv
  |\tilde{\phi}(\mathbf{p})|^2 |g(\mathbf{p})|^2
  \label{eq:sfavg}
\end{align}
contains the momentum-space density $\tilde{\phi}$ as a factor. This provides
a way to extract the momentum-space density of the initial state from the
measured energy-resolved spectra. In the limit of an infinite pulse, the
integral even simplifies to an expression including generalized
Bessel-functions \cite{sfa:fais73}.

\end{document}